Orthogonalized smoothing for rescaled spike and slab models
=======

**Hemant Ishwaran**[*1] **and Ariadni Papana**[2]

*Cleveland Clinic and Case Western Reserve University*

**Abstract:** Rescaled spike and slab models are a new Bayesian variable selection method for linear regression models. In high dimensional orthogonal settings such models have been shown to possess optimal model selection properties. We review background theory and discuss applications of rescaled spike and slab models to prediction problems involving orthogonal polynomials. We first consider global smoothing and discuss potential weaknesses. Some of these deficiencies are remedied by using local regression. The local regression approach relies on an intimate connection between local weighted regression and weighted generalized ridge regression. An important implication is that one can trace the effective degrees of freedom of a curve as a way to visualize and classify curvature. Several motivating examples are presented.

## Contents

## 1. Introduction

Rescaled spike and slab models were introduced in [1] as a Bayesian variable selection method in linear regression models. Such models were shown to possess a

---

*Supported by the NSF Grant DMS-04-05675.

[1]Department of Quantitative Health Sciences Wb4, Cleveland Clinic, 9500 Euclid Avenue, Cleveland, Ohio 44195, USA, e-mail: hemant.ishwarant@gmail.com

[2]Department of Statistics, Case Western Reserve University, 10900 Euclid Avenue, Cleveland, Ohio 44106, USA, e-mail: ariadni.papana@case.edu

*selective shrinkage* property in orthogonal models. This property allows the posterior mean for the coefficients to shrink to zero for truly zero coefficients while for the non-zero coefficients posterior estimates are similar to the ordinary least squares (OLS) estimates. In [2], rescaled spike and slab models were used to analyze multi-group microarray data (an extension of previous work [3]). Selective shrinkage was shown to be a sufficient condition for oracle-like total misclassification. A finite sample adaptive method for selecting variables using this principle was given.

In this manuscript we extend the application of rescaled spike and slab models to smoothing problems. Given an outcome value $Y$ related to a variable $x$ through an unknown function $f(x)$, we would like accurate recovery of $f(x)$. Smoothing is a prediction problem, and an important contribution of the paper is advancing applications of rescaled spike and slab models to prediction settings. However this does not mean selective shrinkage, a core ingredient to model selection, is not at play in a prediction paradigm. Indeed, as shown, selective shrinkage plays a crucial role in adaptive selection of over-parameterized basis functions in response to curvature of $f(x)$.

We consider global smoothing via orthogonal polynomial regression as well as local regression using orthogonal polynomials. Orthogonality is a key ingredient in our approach. Not only does it allow us to exploit the selective shrinkage property of rescaled spike and slab models, which follow as a consequence of orthogonality, but it also greatly improves the computational efficiency of our algorithms. While much work has been done in the area of smoothing, we note there are novel features in our approach potentially useful in applied settings. One important feature being that selective shrinkage allows for greater adaptivity to curvature and greater robustness to misspecification of dimension of basis functions. Secondly, in local regression settings, adaptivity via selective shrinkage can be interpreted in terms of dimensionality and curvature. From this we provide an effective degrees of freedom plot for graphing estimated dimensionality of $f(x)$ as a function of $x$. Such plots provide a simple and powerful way to register curves. Several applications are provided as illustration.

## 2. Rescaled spike and slab models

We begin by first reviewing background theory for rescaled spike and slab models. The underlying setting is the linear regression model where $Y_1, \ldots, Y_n$ are independent responses such that

$$(2.1) \qquad Y_i = \beta_1 x_{i,1} + \cdots + \beta_d x_{i,d} + \varepsilon_i = \mathbf{x}_i^t \boldsymbol{\beta} + \varepsilon_i, \qquad i = 1, \ldots, n.$$

Here $\mathbf{x}_1, \ldots, \mathbf{x}_n$ are non-random (fixed design) $d$-dimensional covariates and $\boldsymbol{\beta} = (\beta_1, \ldots, \beta_d)$ is the unknown coefficient vector. The $\varepsilon_i$ are independent random variables (but not necessarily identically distributed) such that $E(\varepsilon_i) = 0$, $E(\varepsilon_i^2) = \sigma_0^2$ and $E(\varepsilon_i^4) \leq M$ for some $M < \infty$ (the last condition is needed to invoke a triangular central limit theorem later, but is not crucial and can certainly be relaxed). The variance $\sigma_0^2 > 0$ is assumed to be unknown. Throughout we assume $\mathbf{x}_i$ are standardized so that $\sum_{i=1}^n x_{i,k} = 0$ and $\sum_{i=1}^n x_{i,k}^2 = n$ for $k = 1, \ldots, d$ (without loss of generality we assume that there is no intercept term in (2.1)). We shall also assume throughout that $\mathbf{X}$, the $n \times d$ design matrix, is orthogonal, i.e., $\mathbf{X}^t \mathbf{X} = n\mathbf{I}$. As mentioned in the Introduction, this will allow us to exploit certain elegant theories for rescaled spike and slab methods, although, of course, the spike and slab method works for general design matrices.

Spike and slab methods first appeared in the works of [4] and [5] for subset selection in linear regression models. The expression "spike and slab," coined by Mitchell and Beauchamp in [5], referred to the prior for the regression coefficients used in their hierarchical formulation. This was chosen so that the coefficients were mutually independent with a two-point mixture distribution made up of a uniform flat distribution (the slab) and a degenerate distribution at zero (the spike). In [6] a different type of prior was used. This involved a scale mixture of two normal distributions. In particular, the use of a normal prior was highly advantageous and led to a Gibbs sampling method that highly popularized the spike and slab approach; see [7, 8, 9, 10, 11].

As pointed out in [1], priors involving a normal scale mixture distribution, of which [6] is a special example, constitute a wide class of models termed "spike and slab models." A modified class of spike and slab models called "rescaled spike and slab models" was introduced [1]. These new models differed in that the original $Y_i$ values were replaced by new values scaled by the square root of the sample size and divided by the square root of an estimate for $\sigma_0^2$. Rescaling was shown to induce a non-vanishing penalization effect for the posterior mean, and when used in tandem with a continuous bimodal prior, the resulting posterior mean was shown to possess a selective shrinkage property in orthogonal models [1].

A *rescaled spike and slab model* was defined in [1] to denote a Bayesian hierarchical model specified as follows:

$$
\begin{aligned}
(Y_i^* | \mathbf{x}_i, \boldsymbol{\beta}) &\stackrel{\text{ind}}{\sim} \text{N}(\mathbf{x}_i^t \boldsymbol{\beta}, n), \qquad i = 1, \ldots, n, \\
(\boldsymbol{\beta} | \boldsymbol{\gamma}) &\sim \text{N}(\mathbf{0}, \boldsymbol{\Gamma}), \\
\boldsymbol{\gamma} &\sim \pi(d\boldsymbol{\gamma}).
\end{aligned}
$$

(2.2)

Here $Y_i^*$ are the rescaled $Y_i$ values defined by $Y_i^* = \hat{\sigma}^{-1} n^{1/2} Y_i$, where $\hat{\sigma}^2 = ||\mathbf{Y} - \mathbf{X}\hat{\boldsymbol{\beta}}_0||^2 / (n-d)$ is the unbiased estimator for $\sigma_0^2$ based on the full model, and $\hat{\boldsymbol{\beta}}_0$ is the OLS estimate for $\boldsymbol{\beta}$ (other estimators for $\sigma_0^2$ are also possible; these details, however, play a minor role). The value of $n$ used in the first level of the hierarchy in (2.2) is a variance inflation factor introduced to compensate for the rescaling. Moreover, inclusion of $n$ in the hierarchy was shown in [1] to be necessary for selective shrinkage to take place. Without rescaling, shrinkage for the posterior mean vanishes in the limit due to the prior becoming swamped by the likelihood [1].

In (2.2), $\mathbf{0}$ denotes a $d$-dimensional zero vector, $\boldsymbol{\Gamma} = \text{diag}(\gamma_1, \ldots, \gamma_d)$ is a $d \times d$ diagonal matrix and $\pi$ is the prior measure for $\boldsymbol{\gamma} = (\gamma_1, \ldots, \gamma_d)^t$. A Bayesian parameter $\sigma^2$ can also be introduced in (2.2) at the first level of the hierarchy. However, we avoid this approach here and opt for the simpler set up ((2.2)). The rationale for this is the following: (i) we have already removed the effect of $\sigma_0^2$ when rescaling $Y_i$, and (ii) the simpler setup enforces a sparse solution for the posterior mean in ill-determined settings when $d$ is of the same size, or larger, than $n$. Point (ii) is especially relevant as this is the setting we are interested in here.

## 2.1. *Rescaling, the choice of $\pi$, and implications for shrinkage*

In addition to rescaling the response, the prior for $\gamma_k$ must satisfy certain requirements in order for selective shrinkage to occur. A sufficient condition requires the prior to have a bimodal property such that the right tail of the distribution is continuous and such that there is a spike in the distribution near zero (see Theorem 6 of [1] for precise details). One such example is the continuous bimodal prior used in [1, 2, 3]. This prior is induced by a parameterization involving a binary variable and a positive variable with an inverse-gamma distribution. More precisely, define $\gamma_k$ by $\gamma_k = I_k \tau_k^2$, where $I_k$ and $\tau_k^2$ are parameters with priors specified according to

$$
\begin{aligned}
(I_k|v_0, w) & \overset{\text{iid}}{\sim} & (1-w)\,\delta_{v_0}(\cdot) + w\,\delta_1(\cdot), \qquad k = 1, \dots, d, \\
(\tau_k^{-2}|a_1, a_2) & \overset{\text{iid}}{\sim} & \text{Gamma}(a_1, a_2), \\
w & \sim & \text{Uniform}[0, 1].
\end{aligned}
$$

(2.3)

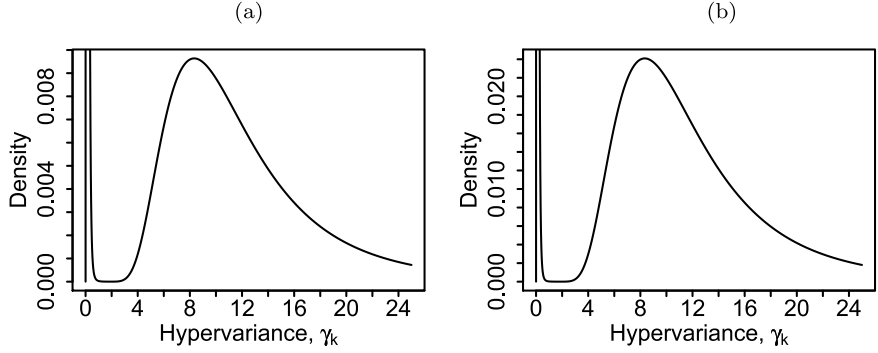

Fig 1. *Conditional density for $\gamma_k$ given $w$: (a) $w = 0.1$, (b) $w = 0.25$. Observe that only the densities height changes as $w$ is varied. One can think of $w$ as a complexity parameter controlling model dimension. Prior based on hyperparameters $a_1 = 5, a_2 = 50$ and $v_0 = 0.005$ as in [1, 2, 3].*

The choice for $v_0$ (a small positive value) and $a_1$ and $a_2$ (the shape and scale parameters for a gamma density) are selected so that $\gamma_k$ has a continuous bimodal distribution with a spike at $v_0$ and a right continuous tail (see Figure 1). Such a prior allows the posterior to shrink a coefficient to zero depending upon the value for $\gamma_k$. Small values heavily shrink a coefficient towards zero.

## 2.2. Selective shrinkage recast in terms of penalization

One can view the posterior mean as a solution to a constrained least squares optimization problem in which the hypervariances are related to penalty terms. This provides us with another way to think about the effects of selective shrinkage. As before, we consider the orthogonal setting where $\mathbf{X}^t\mathbf{X} = n\mathbf{I}$. Let $V_k = E(\nu_k|\mathbf{Y}^*)$ where $\nu_k = \gamma_k/(1 + \gamma_k)$. For our argument it will be easier to think of penalization in terms of $\hat{\boldsymbol{\beta}} = \hat{\sigma}n^{-1/2}\hat{\boldsymbol{\beta}}^*$, where $\hat{\boldsymbol{\beta}}^* = E(\boldsymbol{\beta}|\mathbf{Y}^*)$ denotes the posterior mean for $\boldsymbol{\beta}$ under our rescaled spike and slab model. It can be shown that

$$
\hat{\boldsymbol{\beta}} = \arg\min_{\boldsymbol{\beta}} \left\{ ||\mathbf{Y} - \mathbf{X}\boldsymbol{\beta}||^2 + n \sum_{k=1}^{d} \frac{1 - V_k}{V_k} \beta_k^2 \right\}.
$$

(2.4)

Observe how each $\beta_k$ coefficient in (2.4) is penalized by a unique value $(1 - V_k)/V_k$. The closer $V_k$ is to 1, the smaller the penalty and the less the shrinkage for $\beta_k$, while the closer $V_k$ is to zero, the larger the penalty, and the more $\beta_k$ is shrunk to zero. It is clear the more adaptive $V_k$ is to the true coefficient value, the more accurate variable selection becomes.

This argument can be formalized by studying the asymptotic behavior of $V_k$. Using a similar argument as in [2], one can show that under the spike and slab

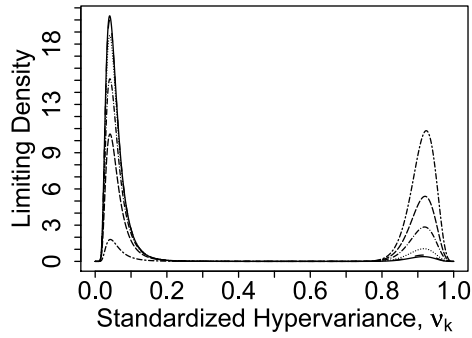

FIG 2. *Limiting density for $\nu_k$ conditioned on $w = 0.1$ and $Z_k^2$ under the null that $\beta_k$ is truly zero. Values for $Z_k^2$ selected from the 25th, 50th, 75th, 90th, 95th and 99th percentiles of a $\chi_1^2$-distribution. Mode on the left increases as $Z_k^2$ decreases, whereas mode on the right decreases.*

model (2.2) with continuous bimodal prior (2.3) (specified in Figure 1), the following holds:

**Theorem 2.1.** *Assume that $\max_{1 \le i \le n} ||\mathbf{x}_i|| / \sqrt{n} \to 0$. If (2.1) represents the true data model, and the coefficient $\beta_k$ for variable $k$ is truly non-zero, then*

$$(2.5) \qquad\qquad V_k \overset{\text{d}}{\rightsquigarrow} 1.$$

*Moreover, if $\beta_k$ is truly zero, then*

$$(2.6) \qquad E(\nu_k | w, \mathbf{Y}^*) \overset{\text{d}}{\rightsquigarrow} \frac{\int_0^1 \nu \exp(\nu Z_k^2 / 2)(1-\nu)^{-3/2} f\big(\nu / (1-\nu) | w\big) \, d\nu}{\int_0^1 \exp(\nu Z_k^2 / 2)(1-\nu)^{-3/2} f\big(\nu / (1-\nu) | w\big) \, d\nu},$$

*where $f(\cdot | w) = (1-w)g_0(\cdot) + w g_1(\cdot)$ is the prior density for $\gamma_{lk}$ given $w$, where $g_0(u) = v_0 u^{-2} g(v_0 u^{-1})$, $g_1(u) = u^{-2} g(u^{-1})$ and*

$$g(u) = \frac{a_2^{a_1}}{(a_1 - 1)!} u^{a_1 - 1} \exp(-a_2 u)$$

*and $Z_k$ has a $N(0, 1)$ distribution.*

Result (2.5) of Theorem 2.1 shows that $V_k$ approaches the value 1 in the case where there is true signal, and hence the penalty for $\beta_k$ in (2.4) vanishes as the sample size increases, and $\beta_k$ will not be shrunk, just as we'd expect and hope for.

Result (2.6) applies to the case when $\beta_k$ is really zero. The term $Z_k$ appearing in (2.6) is the limit of the posterior mean $\hat{\beta}_k^*$ under the null, and thus $Z_k$ reflects the effect of the data on the posterior under the null. In particular, unless $\hat{\beta}_k^*$ is unduly large, the posterior mean for $\nu_k$ should be relatively close to the value under the prior. This has implications for sparse settings. In such cases, the posterior value for $w$ will be small and the posterior for $\nu_k$ conditioned on $w$ (which will look like the prior given $w$) will be concentrated near zero. Thus, the left-hand side of (2.6) should be small and the posterior mean penalized and shrunk towards zero. On the other hand, if $\hat{\beta}_k^*$ is large, then the left-hand side of (2.6) will be large, and there will be less penalization and less shrinkage for $\beta_k$. A large value for $\hat{\beta}_k^*$ is unlikely under the null and in fact is expected only when $\beta_k$ is really non-zero, which is another way to see why (2.5) holds. Figure 2 illustrates how $\nu_k$ might depend upon $\hat{\beta}_k^*$ in a sparse setting under the null that $\beta_k$ is truly zero.

### 3. Orthogonal polynomials: first illustration

For our first illustration we consider a dataset related to spinal bone mineral density (BMD) (see [12] for a more complete description of the data). The response is the relative change in spinal BMD as a function of age in male and female adolescents. Figure 3 plots the results of our analysis. Predicted values for $Y$ based on the posterior of the rescaled spike and slab model (2.2) are superimposed on the figure as solid dark and dashed dark lines for men and women, respectively. The analysis on the left side of the plot is based on an orthogonal polynomial design matrix with $d = 10$ basis functions. Also superimposed are OLS estimates (gray lines).

While the left side of Figure 3 shows some difference between the methods, discrepancies become more apparent if $d$ is allowed to increase. We re-ran the same analysis but using an overly parameterized design involving $d = 25$ basis functions. The right side of Figure 3 records the result. Notice how badly OLS overfits, whereas rescaled spike and slab predictors remain relatively unaffected.

### 3.1. *Comparative analysis using effective kernels*

A more formal comparison between the two approaches can be based on an effective kernel analysis. Effective kernels were introduced in [13] (Chapter 2.8), as a way to evaluate the differences between kernel smoothers. Suppose we have data $(x_i, Y_i)$, $i = 1, \ldots, n$, where $Y_i$ are the response values. It is assumed that

$$(3.1) \qquad\qquad Y_i = f_i + \varepsilon_i, \qquad i = 1, \ldots, n,$$

where $f_i = \mathbf{x}_i^t \boldsymbol{\beta}$ are unknown mean values and $\mathbf{x}_i \in \mathbb{R}^d$ are the values of the pre-chosen underlying $d$ basis functions evaluated at $x_i$. Call a smoother $s(x)$ linear in $Y$, if for each $x_0$

$$s(x_0) = \sum_{j=1}^{n} S_j(x_0) Y_j,$$

where $S_j(x_0)$ depends only upon the $x$-values and not the responses. More generally, let $\mathbf{f} = (f_1, \ldots, f_n)^t$. If $s(x_i)$ is a linear smoother for $f_i$, then

$$\hat{\mathbf{f}} = \mathbf{S}\mathbf{Y}$$

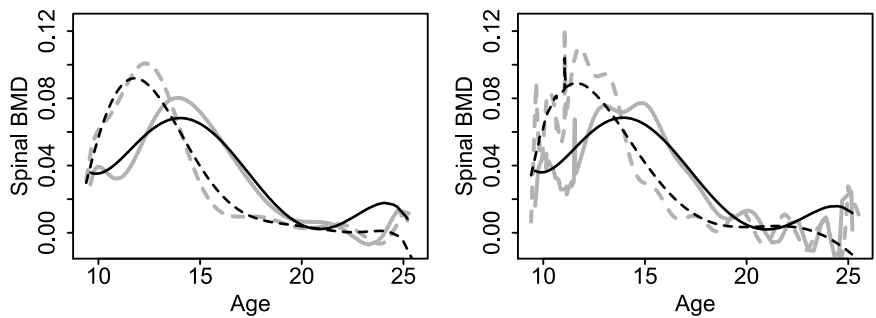

Fig 3. *Left plot: Relative change in spinal BMD as a function of age. Solid dark and dashed dark lines are spike and slab predicted curves for men and women using an orthogonal polynomial design matrix with $d = 10$ basis functions. Gray solid and dashed lines are OLS estimates for men and women. Right plot: Analysis similar as before, but now using an over parameterized basis function, $d = 25$. Note the spiky behavior of the OLS.*

is a linear smoothed estimate of $\mathbf{f}$, where $\mathbf{S}$ is the $n \times n$ *smoother matrix*, $\mathbf{S} = \{s_{i,j}\}$ for $s_{i,j} = S_j(x_i)$. The value $S_i(x_i) = s_{i,i}$ is often referred to as the *effective kernel at* $x_i$ [13, 14]. The effective kernel measures the influence of $x_i$ on $Y_i$. The set of values $\{s_{i,j} : j = 1, \ldots, n\}$, which is the $i$th row of $\mathbf{S}$, is called the effective kernel for $Y_i$. Plotting the effective kernel is a way to compare different smoothers [13].

This idea can be adapted to our setting as follows. First we derive the effective kernel for the OLS estimate. Consider the orthogonal regression setting in which $\mathbf{X}^t\mathbf{X} = n\mathbf{I}$. Let $\mathbf{x}_{(k)}$ denote the $k$th column of the design matrix $\mathbf{X}$. It follows that $\hat{\mathbf{f}} = \mathbf{SY}$, where

$$(3.2) \qquad \mathbf{S} = \mathbf{X}(\mathbf{X}^t\mathbf{X})^{-1}\mathbf{X}^t = n^{-1}\sum_{k=1}^{d}\mathbf{x}_{(k)}\mathbf{x}_{(k)}^t.$$

The effective kernel for $Y_i$ is $n^{-1}\sum_{k=1}^{d} x_{i,k}\mathbf{x}_{(k)}^t$ and the effective kernel at $x_i$ is $s_{i,i} = n^{-1}\mathbf{x}_i^t\mathbf{x}_i$.

The notion of an effective kernel needs to be slightly modified to handle adaptive penalization. We adopt the notion of an adaptive smoother matrix that allows the effective kernel to depend upon both $x_i$ and $Y_i$. Define the spike and slab predictor as $\hat{\mathbf{f}}^* = \mathbf{X}\hat{\boldsymbol{\beta}}$.

**Theorem 3.1.** *Under othogonality, the spike and slab predictor for $Y_i$ in (3.1) can be written as $\hat{\mathbf{f}}^* = \mathbf{S}^*\mathbf{Y}$, where*

$$(3.3) \qquad \mathbf{S}^* = n^{-1}\sum_{k=1}^{d} V_k\mathbf{x}_{(k)}\mathbf{x}_{(k)}^t.$$

*One can conceptualize $\mathbf{S}^*$ as an adaptive linear smoother matrix. Consequently, the effective kernel for $Y_i$ is defined as $n^{-1}\sum_{k=1}^{d} V_k x_{i,k}\mathbf{x}_{(k)}^t$, and the effective kernel at $x_i$ is $s_{i,i}^* = n^{-1}\sum_{k=1}^{d} V_k x_{i,k}^2$.*

Figure 4 shows the effective kernels at $x_i$ for the OLS and rescaled spike and slab predictors, where $x_i$ is age. The plots are based on the over-parameterized orthogonal polynomial design involving $d = 25$ basis functions. The large number of predictors helps to emphasize the non-robustness of the OLS estimate. Note especially how the OLS estimate is affected by the points near the edges of the plots. In contrast, note the robustness of the spike and slab approach.

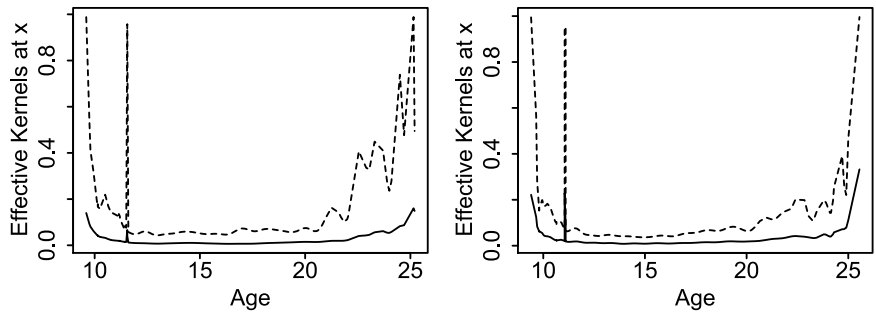

FIG 4. *Left plot: Effective kernels at $x_i$, $i = 1, \ldots, n$, for men using over-parameterized design, $d = 25$ (rescaled spike and slab values depicted by thick line; OLS by dashed line). Right plot: Effective kernels for women.*

### *3.2. Effective degrees of freedom*

The smoother matrix provides information about the nature of a predicted curve. Ideally, however, we would like a rigorous and systematic manner in which to summarize this information as a way to *register* (classify) a curve. One way to rigorously compare curves is to use the notion of the *effective degrees of freedom* [13]. For any smoother matrix $\mathbf{S}$, the effective degrees of freedom, $\mathscr{D}_f$, is defined as

$$\mathscr{D}_f(\mathbf{S}) = \mathrm{tr}(\mathbf{S}) = \sum_{i=1}^{n} s_{i,i}.$$

For the OLS smoother matrix (3.2), $\mathscr{D}_f(\mathbf{S}) = n^{-1} \sum_{i=1}^{n} \sum_{k=1}^{d} x_{i,k}^2 = d$ (the last identity on the right is due to orthogonality). Meanwhile, for the spike and slab smoother matrix (3.3), we have the following corollary to Theorem 3.1.

**Corollary 3.1.** *Under the conditions of Theorem 3.1,*

$$\mathscr{D}_f(\mathbf{S}^*) = n^{-1} \sum_{i=1}^{n} \sum_{k=1}^{d} V_k x_{i,k}^2 = \sum_{k=1}^{d} V_k \leq d.$$

*Hence, the degrees of freedom for the spike and slab smoother is bounded by the dimension of the underlying polynomial basis.*

Observe how $\{V_k\}$, the shrinkage parameters, dictate the degrees of freedom. The larger the value, the more degrees of freedom used up, and the less shrinkage there is. In the analysis presented earlier using a saturated design ($d = 25$), the effective degrees of freedom are 4.2 and 5.8 for men and women, respectively, indicating more overall shrinkage for men and evidence of differences in the two curves.

Effective degrees of freedom are useful for assessing overall differences between curves. However the method is limited in its ability to register a curve, as it reduces the overall properties of a curve to a single summary value. In the next section we illustrate a much more effective way to register curves.

## 4. Local regression

In this section we illustrate how rescaled spike and slab models can be used for local regression, an alternative method of smoothing [15, 16]. By exploiting orthogonality, and by drawing connections to generalized ridge regression, we show that rescaled spike and slab predictors can be viewed as local regression smoothers with a local smoother matrix whose effective degrees of freedom can be traced over $x$ as a way to characterize curvature of the underlying function. Another nice feature of using rescaled spike and slab models, just like in global smoothing, is that we end up being fairly robust to the choice of the dimension of the underlying basis functions.

First let's review some background on local regression. In local regresssion, for a given $x_i$, rather than performing a global regression to estimate $\mathrm{f}_i$, one instead fits a weighted regression model using weighted least-squares, with weights for an observation $x$ chosen by how close they are to $x_i$. This results in a local estimator

$$\hat{\mathbf{f}}(x_i) = (\hat{\mathrm{f}}_{i,1}, \dots, \hat{\mathrm{f}}_{i,n})^t$$

in which the $i$th coordinate, $\hat{\mathrm{f}}_{i,i}$, is used as an estimator for $\mathrm{f}_i = E(Y_i)$. Unlike (3.1), however, the relationship between $\mathrm{f}_i$ and $x_i$ can vary with $i$.

As well known, a local regression predictor is nothing more than a weighted least squares predictor. That is, for a given $x_i$, let $\mathbf{b}_{i,j} = (b_{i,j,1}, \cdots, b_{i,j,d})^t$ be the values of the $d$ basis functions chosen for $x_i$ evaluated at $x_j$. The local regression predictor is defined as $\hat{\mathbf{f}}(x_i) = \mathbf{B}_i \hat{\boldsymbol{\beta}}_W$, where

$$(4.1) \qquad \hat{\boldsymbol{\beta}}_W = \arg\min_{\boldsymbol{\beta}} \left\{ \sum_{j=1}^{n} \left( Y_j - \sum_{k=1}^{d} \beta_k b_{i,j,k} \right)^2 K\left( \frac{x_j - x_i}{h} \right) \right\},$$

and $K(\cdot)$ is a positive kernel function with unknown bandwidth parameter $h > 0$. Solving, it can be shown that $\hat{\mathbf{f}}(x_i)$ is the weighted least squares predictor,

$$(4.2) \qquad \hat{\mathbf{f}}(x_i) = \mathbf{B}_i (\mathbf{B}_i^t \mathbf{W}(x_i) \mathbf{B}_i)^{-1} \mathbf{B}_i^t \mathbf{W}(x_i) \mathbf{Y}$$

where $\mathbf{B}_i$ is the $n \times d$ design matrix with $j$th row $\mathbf{b}_{i,j}$, and $\mathbf{W}(x_i) = \text{diag}\{W_{i,j}\}$ is the $n \times n$ diagonal weight matrix, where

$$W_{i,j} = K\left( \frac{x_j - x_i}{h} \right), \qquad j = 1, \ldots, n.$$

See [14] for details.

**Example 4.1.** A popular basis function expansion for local regression is in terms of polynomials [17, 18]. In this case, the design matrix is

$$(4.3) \qquad \mathbf{B}_i = \begin{pmatrix} 1 & x_1 - x_i & \cdots & (x_1 - x_i)^d \\ 1 & x_2 - x_i & \cdots & (x_2 - x_i)^d \\ \vdots & \vdots & \vdots & \vdots \\ 1 & x_n - x_i & \cdots & (x_n - x_i)^d \end{pmatrix}_{n \times (d+1)}.$$

Note that $\mathbf{B}_i$ has rank $d + 1$ because we always include an intercept term. The rationale for using a polynomial expansion follows by considering an expansion of of $E(Y_j) = f(x_j)$ around $x_i$. Since,

$$f(x_j) = f(x_i) + \sum_{k=1}^{p} \frac{(x_j - x_i)^k}{k!} f^{(k)}(x_i) + R_j,$$

where $R_j$ is a small remainder term, the local weighted regression (4.1) is (approximately) the value for $\boldsymbol{\beta}$ minimizing

$$\sum_{j=1}^{n} \left( f(x_i) + \sum_{k=1}^{p} \frac{(x_j - x_i)^k}{k!} f^{(k)}(x_i) - \left( \beta_0 + \sum_{k=1}^{d} \beta_k (x_j - x_i)^k \right) \right)^2 K\left( \frac{x_j - x_i}{h} \right).$$

If $d = p$, then $\beta_0$ estimates $f(x_i)$ while $\beta_k$ estimates $f^{(k)}(x_i)/k!$ for $k = 1, \ldots, d$. The local regression predictor is

$$\hat{\mathbf{f}}_{i,j} = \hat{\beta}_{0,W} + \sum_{k=1}^{d} \hat{\beta}_{k,W} (x_j - x_i)^k, \qquad j = 1, \ldots, n,$$

which should be a good approximation to $f(x_j)$ when $x_j$ is near $x_i$.

### *4.1. Rescaled spike and slab weighted regression*

The representation (4.2) presents an immediate tie-in to the spike and slab methodology. The rescaled posterior mean, $\hat{\boldsymbol{\beta}}$, from (2.2) is a model averaged generalized ridge regression (GRR) estimator, expressible as

$$\hat{\boldsymbol{\beta}} = E\left\{ \left( \mathbf{X}^t\mathbf{X} + n\boldsymbol{\Gamma}^{-1} \right)^{-1} \mathbf{X}^t\mathbf{Y} \Big| \mathbf{Y}^* \right\}.$$

It is not hard to see that by appropriately introducing a weighting matrix into the hierarchy, that one can arrive at a model averaged weighted GRR estimator, and a smoother of the form (4.2). The advantage of this type of approach is that the resulting smoother will be based on an estimator that uses adaptive penalization.

In this modification, similar to (4.3), we work with a polynomial basis that depends upon $i$. However, our polynomial basis will be strictly orthogonal. Let

$$\mathbb{I}_{i,h} = \left\{ j : K\left( \frac{x_j - x_i}{h} \right) > 0 \right\}.$$

For an orthogonal basis we define $\mathbf{B}_i$ to be the design matrix for $x_i$ obtained using a $d$-degree orthogonal basis using only those $x_j$ values where $j \in \mathbb{I}_{i,h}$.

For each $j \in \mathbb{I}_{i,h}$, define $Y_j^* = \hat{\sigma}_i^{-1} n_i^{1/2} Y_j$, where $n_i$ is the cardinality of $\mathbb{I}_{i,h}$ and $\hat{\sigma}_i^2$ is an estimator for $\sigma_0^2$ for the set of responses, $\{Y_j : j \in \mathbb{I}_{i,h}\}$. We use the estimator due to [19],

$$\hat{\sigma}_i^2 = \frac{1}{2(n_i - 1)} \sum_{j=1}^{n_i - 1} (Y_{(j+1)} - Y_{(j)})^2,$$

where $Y_{(j)}$ is the $Y$-value corresponding to the $j$th ordered $x$-value in $\mathbb{I}_{i,h}$ (the estimator is most easily computed by sorting the $x$ values).

Let $\mathbf{Y}_i^*$ be the vector of the rescaled values $Y_j^* = \hat{\sigma}_i^{-1} n_i^{1/2} Y_j$ for $j \in \mathbb{I}_{i,h}$. Let $\mathbf{W}_i$ be the subset of $\mathbf{W}(x_i)$ corresponding to those $j \in \mathbb{I}_{i,h}$. For a given $x_i$, the modified rescaled spike and slab model is

$$
\begin{aligned}
(\mathbf{Y}_i^* | \mathbf{B}_i, \mathbf{W}_i, \boldsymbol{\beta}) &\sim& \mathrm{N}(\mathbf{B}_i\boldsymbol{\beta}, n_i\mathbf{W}_i^{-1}), \\
(\boldsymbol{\beta} | \boldsymbol{\gamma}) &\sim& \mathrm{N}(\mathbf{0}, \boldsymbol{\Gamma}), \\
\boldsymbol{\gamma} &\sim& \pi(d\boldsymbol{\gamma}).
\end{aligned}
$$

(4.4)

Consider the following theorem which characterizes the spike and slab predictor $\hat{\mathbf{f}}^*(x_i) = \mathbf{B}_i\hat{\boldsymbol{\beta}}_{i,W}$, where $\hat{\boldsymbol{\beta}}_{i,W}$ is the rescaled posterior mean from (4.4). We use this result later to explicitly characterize the smoother matrix and its effective degrees of freedom under orthogonality.

**Theorem 4.1.** *Under the Bayesian hierarchy (4.4), the spike and slab local predictor can be expressed as*

$$\hat{\mathbf{f}}^*(x_i) = (\hat{\mathbf{f}}_{i,1}^*, \dots, \hat{\mathbf{f}}_{i,n}^*)^t = \mathbf{S}_{i,h}^*\mathbf{Y}_i,$$

*where $\mathbf{S}_{i,h}^*$ is the model averaged smoothing matrix defined by*

$$\mathbf{S}_{i,h}^* = \mathbb{E}\left\{ \mathbf{B}_i \left( \mathbf{B}_i^t\mathbf{W}_i\mathbf{B}_i + n_i\boldsymbol{\Gamma}^{-1} \right)^{-1} \mathbf{B}_i^t\mathbf{W}_i \Big| \mathbf{Y}_i^* \right\}.$$

*Note that the smoother matrix $\mathbf{S}_{i,h}^*$, unlike (4.2), takes advantage of adaptive penalization.*

### 4.2. Orthogonality

Our construction for the basis ensures that $\mathbf{B}_i^t\mathbf{B}_i = n_i\mathbf{I}$. However, in order to fully exploit orthogonality, we additionally require that

$$\mathbf{B}_i^t\mathbf{W}_i\mathbf{B}_i = n_i\mathbf{I}. \tag{4.5}$$

For (4.5) to hold we must have $\mathbf{W}_i = \mathbf{I}$. The simplest way to satisfy this condition is to use a nearest neighbour kernel. For a fixed bandwidth value $h$, let

$$K\left(\frac{x}{h}\right) = 1\{|x| < h\}.$$

The nearest neighbour kernel puts a weight of 1 on all values of $x$ within a distance of $h$ to zero. Using such a kernel implies that $\mathbf{W}_i = \mathbf{I}$ and $\mathbb{I}_{i,h} = \{j : |x_j - x_i| < h\}$.

Shrinkage, just as in the global orthogonal regression setting, is intimately related to the degrees of freedom of the smoother matrix. Consider the following corollary to Theorem (4.1) characterizing effective degrees of freedom under orthogonality.

**Corollary 4.1.** *Under the orthogonality assumption (4.5), the local smoother matrix for the rescaled spike and slab predictor is* $\mathbf{S}_{i,h}^* = n_i^{-1}\mathbf{B}_i\mathbf{V}_i\mathbf{B}_i^t$. *The effective degrees of freedom of* $\mathbf{S}_{i,h}^*$ *equals*

$$\mathscr{D}_f(\mathbf{S}_{i,h}^*) = n_i^{-1}tr(\mathbf{B}_i\mathbf{V}_i\mathbf{B}_i^t) = n_i^{-1}tr(\mathbf{B}_i^t\mathbf{B}_i\mathbf{V}_i) = \sum_{k=1}^{d} V_{i,k} \leq d,$$

*where* $\mathbf{V}_i = diag\{V_{i,k}\}$ *and*

$$V_{i,k} = \mathbb{E}\left(\frac{\gamma_k}{1+\gamma_k}\Big|\mathbf{Y}_i^*\right), \qquad k = 1, \ldots, d.$$

*Hence, the degrees of freedom of the spike and slab local smoother is bounded by the dimension of the local polynomial basis.*

The effective degrees of freedom can be used to provide insight into the geometry of a curve $f$. If the effective degrees of freedom is large, $f$ will possess higher order local curvature, whereas if the degrees of freedom are small, $f$ is likely to be flat. Plotting $\mathscr{D}_f(\mathbf{S}_{i,h}^*)$ is therefore a way to register a curve and to identify key differences between curves. We illustrate this concept by way of three different examples.

### 4.3. Spinal BMD data revisited

For our first example, we applied the local rescaled spike and slab model (4.4) to the previously analyzed BMD data. For the analysis, we used a nearest neighbour kernel with a bandwidth set at $h = 1$, corresponding to one year of age. For the basis function we used cubic orthogonal polynomials. Figure 5 plots the spike and slab predictor for men and women (line types as in Figure 3). Also superimposed on the figure are predicted curves using Friedman's supersmoother (implemented in the programming language R by the call "supsmu(x,y)"). The two methods agree closely, although Friedman's smoother appears to over-smooth the data for men. The right-hand side of Figure 5 plots the effective degrees of freedom for men and women. One can immediately see a phase shift in the figure, signifying distinct modes for the two curves. Also, overall, there is significantly less shrinkage for women with degrees of freedom being positive over a much wider region than men. In both cases, curves eventually flatten out at around age 20.

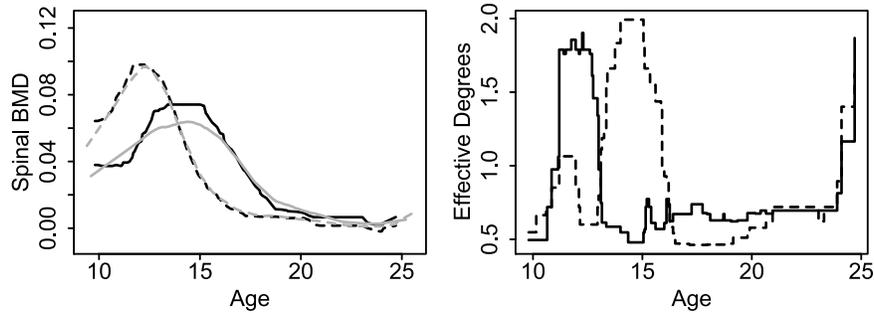

Fig 5. *Left plot: Local kernel regression for BMD data via rescaled spike and slab models with orthogonal polynomial cubic basis functions. Solid dark and dashed dark lines are spike and slab predictors for men and women, respectively. Gray solid and dashed lines are Friedman's super-smoother for men and women. Right plot: Effective degrees of freedom of spike and slab local smoother (solid lines are men, dashed lines are women).*

### 4.4. Cosmic microwave background radiation

As another illustration we look at data related to cosmic microwave background (CMB) radiation [20]. Here, the value for $x$ is the multipole moment and $Y$ is the estimated power spectrum of the temperature fluctuations. The outcome is sound waves in the cosmic microwave background radiation, which is the heat left over from the big bang.

We used the same strategy and settings as before. For the bandwidth we used $h = 25$ which was estimated prior to fitting using generalized cross-validation. Results from the analysis are depicted in Figure 6 with plots zoomed in on different regions of $x$ in order to help visualize the varying curvature. The bottom right plot of Figure 6 shows the effective degrees of freedom. The plot suggests the presence of at least 4 distinct inflection points. In particular, note that initially for $x < 200$ there is a steep increase in the curve signified by the effective degrees of freedom being roughly constant at 3.0. At around $x = 200$ there is a significant drop in the effective degrees of freedom, followed by an increase and a flattening out until around $x = 400$. The drop at $x = 200$ indicates the first inflection point. At $x = 400$ there is another drop in the effective degrees of freedom. Similarly, there is a drop near $x = 600$ and $x = 800$. All told, this suggests at least 4 distinct inflections, all appearing in multiples of 200 starting at $x = 200$.

### 4.5. Mass spectrometry protein data

The study of proteins is critical to understanding living organisms at the molecular level as proteins are the main components of physiological pathways of cells. Proteomics, the study of proteins on a large scale, is often considered the natural step after genomics in the study of biological systems. Greatly complicating any system-wide analysis of proteins, however, is the dynamic nature of the proteome, which constantly changes through its biochemical interactions with the genome and the environment. While the challenges faced by proteomics are great, the benefits at the same time are potentially huge. For example, by studying protein differences for diseased individuals, one might be able to discover pathways responsible for these differences, which in turn could lead to novel biomarkers for identifying disease.

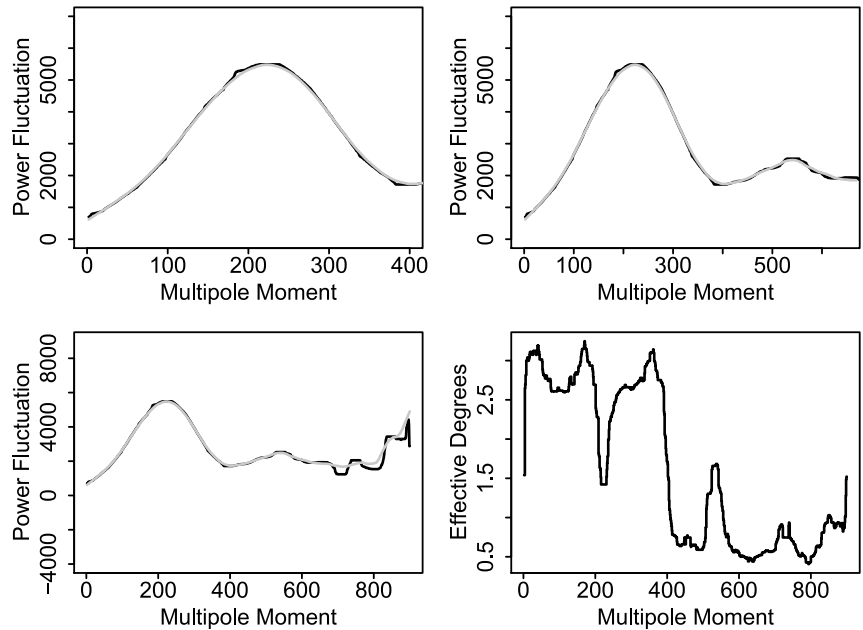

Fig 6. *First 3 plots (top to bottom left to right) are rescaled spike and slab local predictors (thick dark lines) for CMB data. Gray lines are Friedman's smoother. Bottom right plot: Effective degrees of freedom for rescaled spike and slab predictor. Note the presence of 4 modes suggested by this last plot.*

One promising technology for profiling protein behavior is SELDI-TOF-MS (surface enhanced laser desorption/ionization time-of-flight mass spectrometry). In this technology, homogeneous biological samples are placed on the active surface of an array. The protein samples are washed and an energy absorbing molecule solution is placed on the surface of the array and allowed to crystalize. The array is then queried by a laser which ionizes the proteins in the sample. Charged gaseous peptides are emitted and their intensity is detected downstream. The mass over charge ratio ($m/z$) of a peptide-ion is determined from the recorded TOF (time-of-flight). The data collected from a SELDI-TOF-MS experiment consists of the intensity (abundance) of proteins in the sample for a given $m/z$ ratio. One can think of the set of these two values as constituting a spectra. Each biological sample produces one spectra and it is of interest to study differences in spectra as a function of phenotype. See [21] for more details and further references.

Identification of unique peaks in the spectra, a method commonly referred to as peak identification, is a crucial part of analyzing mass spectrometry data. From a statistical perspective, peak identification can be recast as a smoothing problem where the goal is to identify modes in the data after appropriate smoothing. The outcomes are the spectrometry intensity measurements, whereas $x$ is the specific $m/z$ ratio. To illustrate how our spike and slab method can be used for peak detection, we analyzed a set of 8 calibration spectra available as part of the "PROcess" library [21] in the Bioconductor R-suite. The data is unique because it is known a priori that the same 5 proteins are present in each of the 8 samples.

The results from the analysis are plotted in Figure 7 (we note that the data was first baseline normalized prior to analysis). The black lines in the top plot are the rescaled spike and slab smoothed predictors for each spectra. We used orthogonal

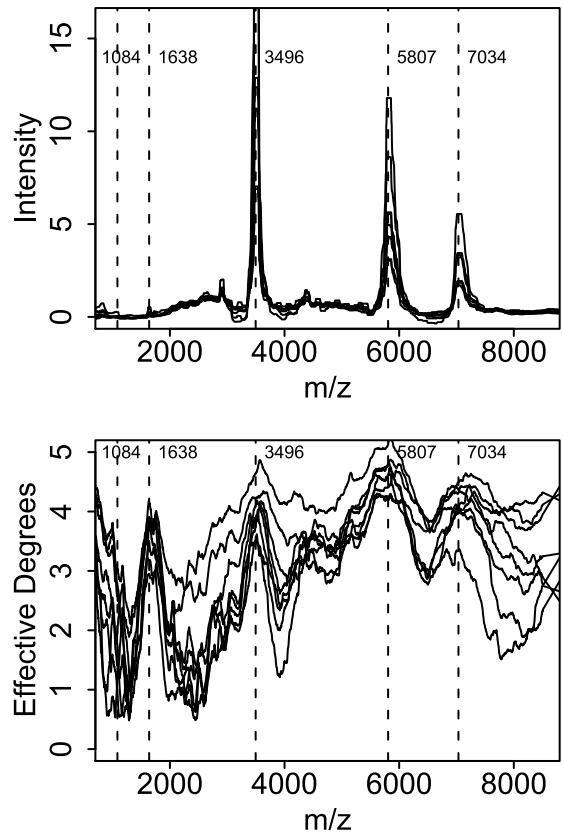

F<small>IG</small> 7. *Mass spectrometry calibration data: 8 spectra, each comprising 13,468 distinct m/z ratios (horizontal axis constrained to a subset of observed m/z ratios to help zoom in figure). Top plot: Solid lines are rescaled spike and slab predictors and dashed vertical lines indicate known unique proteins. Bottom plot: Effective degrees of freedom.*

polynomials of degree 5 (the high degrees of freedom used due to the spiky nature of the data). The bandwidth was set at $h = 50$. Superimposed are 5 dashed vertical lines indicating the 5 distinct proteins. Interestingly, we find that 3 of the 5 proteins are clearly identified in all 8 spectra. However, the two smallest proteins $m/z = 1084$ and $m/z = 1638$ are less visible, the protein at $m/z = 1084$ especially so. There is also evidence of at least 2 additional peaks at approximately $m/z = 3500$ and $m/z = 4500$. The effective degrees of freedom plot, also given in Figure 7, confirms these findings. The plot also indicates that overlap of spectra is sub-par suggesting further normalization of the data is needed.

**Acknowledgments.** The authors are grateful to Bertrand Clarke and the referee who reviewed an earlier version of the paper. We also thank Liang Li for his comments.